\documentclass[12pt,a4paper,reqno]{article}

\usepackage{amsmath,latexsym,amssymb,amsthm}
\usepackage{graphicx}

\newtheorem{thm}{Theorem}

\newcommand{\trans}{^{\mathrm{T}}}
\newcommand{\up}{\uparrow}
\newcommand{\dn}{\downarrow}

\begin{document}

\title{\bf{Local symmetry properties of pure 3-qubit states.}}

\author{H. A. Carteret$^1$ and A. Sudbery$^2$ \\[10pt] \small Department of Mathematics,
University of York, \\[-2pt] \small Heslington, York, England YO10 5DD\\
\small $^1$Email: hac100@york.ac.uk \\ \small $^2$Email: as2@york.ac.uk}

\date{25 January 2000, revised 20 June 2000}

\maketitle

\begin{abstract}
\item
Entanglement types of pure states of three spin-$\tfrac{1}{2}$ particles
are classified by means of their stabilisers in the group of local
unitary transformations.  It is shown that the stabiliser is generically
discrete, and that a larger stabiliser indicates a stationary value for
some local invariant.  We describe all the exceptional states with
enlarged stabilisers.
\end{abstract}

\section{Introduction}
  
It is only relatively recently that the importance of entanglement has
been fully realised.  Not only, as Schr\"{o}dinger emphasised
\cite{schrocat}, 
does it constitute one of the chief differences between
classical and quantum mechanics, and the main obstacle to an intuitive
understanding of quantum mechanics; the recent discovery 
is that it is also a resource, yielding much greater capabilities than
classical physics in information processing and communication (see for
example \cite{LoPS}).

It is therefore important to analyse and measure this resource.  A full
analysis has so far been achieved only for pure state systems with two
component parts \cite{poproh,monoton}; for multipartite systems there
are several different possible measures of entanglement
\cite{thap,eameasmpe,tangle,Grassl,H3LimEM,contBE}, the relation between
them being incompletely understood.  A full quantitative analysis of
entanglement even for pure states of three-part systems appears to be
difficult (but see \cite{Grassl}).  Our aim in this paper is to give a
\emph{qualitative} analysis of the entanglement of such states, using
group-theoretic methods to classify the possible kinds of entanglement.

The nature of the entanglement between the parts of a composite system
should not depend on the labelling of the basis states of each of the
part-systems; it is therefore invariant under unitary transformations of
the individual state spaces. Such transformations are referred to as
\emph{local} unitary transformations, though there is no implication
that the part-systems should be spatially separated.  If the
part-systems have individual state spaces $\mathcal{H}_1, \ldots,
\mathcal{H}_n,$ so that the space of pure states of the composite
system is $\mathcal{H}_1 \otimes \ldots \otimes \mathcal{H}_n,$ then a
local unitary transformation is of the form $U_1 \otimes ... \otimes
U_n$ where $U_i$ is a unitary operator on $\mathcal{H}_i.$ The set of
all such transformations is a group $G,$ whose orbits in
$\mathcal{H}_1 \otimes ... \otimes \mathcal{H}_n$ are equivalence
classes of states with the same entanglement properties. Each orbit
therefore corresponds to a complete specification of entanglement.  The
orbits can be classified by their dimensions, which are determined by
the stabiliser subgroups of points on the orbit; the relation is
\begin{equation}
 \text{dim}\mathcal{O} + \text{dim}\mathcal{S} = \text{dim}G
\end{equation}
where $\mathcal{O}$ is an orbit and $\mathcal{S}$ is the stabiliser of
any point on $\mathcal{O},$ i.e. the set of elements of $G$
which leave a point unchanged (different points on the same orbit have
conjugate stabilisers, which have the same dimension).

This paper is concerned with pure states of three spin-$\tfrac{1}{2}$
particles $(n=3;$ $\mathcal{H}_1=\mathcal{H}_2=\mathcal{H}_3=\mathbb{C}^2.)$ 
We will show that for most states (all but a set of lower dimension) the
stabiliser is discrete, so the dimension of the orbit is the same as
that of the group $G.$  Classifying types of entanglement by
the dimension of the orbit is therefore equivalent to identifying
certain exceptional types of entanglement, which can be expected to be
particularly interesting and important.  One way in which this manifests
itself is that any such exceptional entanglement is necessarily
associated with an extreme value of one of the local invariants which
form coordinates in the space of entanglement types, and from which any
measure of entanglement must be constructed.

The organisation of the paper is as follows.  In Section 2 we review the
case of two spin-$\tfrac{1}{2}$ particles.  The results here are well-known, but
we include them for the sake of completeness and orientation.  In
Section 3 we prove the general theorem about three spin-$\tfrac{1}{2}$
particles mentioned in the preceding paragraph.  Section 4 consists of
the theorem concerning the association between enlarged stabilisers and
stationary values of invariants.  Section 5 contains the
classification of exceptional entanglement types in the system of three
spin-$\tfrac{1}{2}$ particles, in which we examine all the states which
are identified as non-generic in the theorem of Section 3. Section 6 is
a summary listing these exceptional states.  They are illustrated by
means of plots of their two-particle entanglement entropies in an
appendix.

\paragraph{Acknowledgments}

We are indebted to Dr. Ian McIntosh for a helpful conversation, and to
Prof. A. Popov for drawing reference \cite{maxref} to our attention.

\medskip

\thanks{The research of the first author was supported by the EPSRC.}

\section{The Stabiliser for the 2-particle case.}

A pure state of two spin-$\tfrac{1}{2}$ particles can be written as 
\begin{equation}
 | \Psi \rangle = \sum_i t_{ij} |\psi_i \rangle | \psi_j \rangle
\end{equation}
where $\{ | \psi_1 \rangle , | \psi_2 \rangle \}$ is a basis of
one-particle states.  Having fixed this basis, we can identify the state
$|\Psi \rangle $ with the matrix of coefficients $T= (t_{ij}).$  The group
of local transformations is 
\begin{equation}
 G_2 = U(1) \times SU(2) \times SU(2), 
\end{equation}
since the phases in the individual unitary transformations can be
collected together.  The effect of a local transformation
$(e^{i\theta},X,Y)$ on $T$ is to change it to $e^{i\theta}XTY\trans,$ so the
condition for $(e^{i\theta},X,Y)$ to belong to the stabiliser of
$|\Psi\rangle$ is 
\begin{equation}
 T = e^{i\theta} XTY \trans.
\end{equation}
For a 2-particle state, we can always perform a Schmidt
decomposition, so we need only consider states for which
\begin{equation*}
  T =
  \begin{pmatrix} p & 0 \\
                  0 & q 
  \end{pmatrix},
\end{equation*} 
i.e.
\begin{equation*}|\Psi \rangle = p|\up \rangle|\up \rangle + q|\dn \rangle|\dn \rangle
\end{equation*} 
where $p,q$ are real and positive.  Multiplying the stabiliser equation
on the right by $\overline{Y},$ where the overbar denotes complex
conjugation, and writing
\begin{align}
 &X = 
 \begin{pmatrix} r & s \\
                 -\overline{s} & \overline{r}
 \end{pmatrix}, \quad
 &Y = 
 \begin{pmatrix} g & h \\
                 -\overline{h} & \overline{g}
 \end{pmatrix},
\end{align}
we obtain:
\begin{equation*}
  \begin{pmatrix} p & 0 \\
                  0 & q
  \end{pmatrix}
  \begin{pmatrix} \overline{r} & \overline{s} \\
                  -s      & r
  \end{pmatrix}             
  = e^{i\varphi}  
  \begin{pmatrix}   g      &  h \\
                  -\overline{h} & \overline{g} 
  \end{pmatrix}
  \begin{pmatrix} p & 0 \\
                  0 & q 
  \end{pmatrix}. 
\end{equation*}
For given $p,q,$ we want to find the set of solutions $(\varphi,g,h,r,s)$
with $\varphi$ real and $g,h,r,s \in \mathbb{C},$ with $|g|^2 + |h|^2 = 1, |r|^2 +
|s|^2=1.$  If $p \neq 0$, then  $g = \overline{r} e^{-i\varphi}$.  If $q \neq 0$,
then $\overline{g} = r e^{-i\varphi}$.  Therefore either $r = 0$ or
$\varphi = n \pi$.  Also 
\begin{equation*}
h = \frac{p}{q}\overline{s}e^{-i\varphi} = \frac{q}{p}\overline{s}e^{i\varphi} 
\end{equation*}
So unless $p = q$ (since $p$ and $q$ were obtained by a Schmidt
decomposition, they cannot be negative) we must have
$\frac{p}{q} e^{-i\varphi}-\frac{q}{p} e^{i\varphi} \neq 0$
and so $s = 0$.  The states now fall naturally into three classes:

\paragraph{Case 1: The General case.}

If $p \neq 0$ and $q \neq 0$ and $p \neq q$ then $s = h = 0$ and
$e^{i\varphi} = e^{-i\varphi} = \pm 1$ so we can absorb that external sign into
$X.$  This is the subgroup 
\begin{align}\label{n2gen}
  &\varphi = 0, \quad &X = e^{i\nu\sigma_z}, \quad \quad \quad &Y = \bar{X}.
\end{align}
The stabiliser has one parameter, $\nu$.

\paragraph{Case 2: The Unentangled case.}

Without loss of generality, we can take $p = 1$, $q = 0$.  Putting
$g = e^{i\theta}$, this is the subgroup
\begin{equation}\label{n2unent}
(e^{i\varphi},g,r) = (e^{i\varphi},e^{i\theta},e^{-i(\varphi + \theta)})
\end{equation} 
The stabiliser has two parameters, $\varphi$ and $\theta$.

\paragraph{Case 3: The Maximally Entangled Case.}

This occurs when $p = q = 1/\sqrt 2$.  Then 
\begin{align*}
  g &= \overline{r} e^{-i\varphi} = \overline{r} e^{i\varphi} \\
  h &= -s e^{-i\varphi} = -s e^{i\varphi}
\end{align*}
So $\varphi = n \pi$ (or else we'd have to have $r = s = 0$ which is
impossible).  Thus $g = \pm \overline{r}$ and $h = \mp s$, giving the
three-parameter subgroup defined by $Y= \pm \overline{X},$ where $X$    
can be anything in $SU(2).$  

These results illustrate how the occurrence of a state with special
physical significance is signalled by a change in the stabiliser. In
Case 2 above the states are factorisable, so there is minimal
entanglement: the stabiliser increases from one- to two-dimensional. In 
Case 3, on the other hand, the entanglement is maximal as measured by
the entropy of entanglement
\[
  S = p^2 \ln p^2 + q^2 \ln q^2
\]
or equivalently by the 2-tangle \cite{tangle}
\[
  \tau = p^2 q^2 = p^2(1-p^2)
\]
(see Section 6). We note that the stabiliser for these states is even
larger, being three-dimensional. 

This association between an enlarged stabiliser and a maximum or minimum
of an invariant measure of entanglement is a general phenomenon, as will
be proved in Section 4.

\section{The 3 spin-$\tfrac{1}{2}$ Particle Generic Stabiliser.}

In this section we will show that the generic pure state of three
spin-$\tfrac{1}{2}$ particles has a discrete stabiliser in the group 
\begin{equation}
 G_3 = U(1) \times SU(2) \times SU(2) \times SU(2)
\end{equation}
of local unitary transformations.  This is in contrast to the case of two
particles, where, as shown in the previous section, every state has a
stabiliser which is at least one-dimensional. In the course of the proof we will
identify those exceptional states for which the stabiliser might have 
dimension greater than zero.  For ease of later reference, we will label 
those steps in the argument whose failure could produce such nongeneric
behaviour. 

\begin{thm}\label{tony's} 
Let $|\Psi\rangle$ be a pure state of three spin-$\tfrac{1}{2}$
particles, and let $L(\Psi)$ be the Lie algebra of the
stabiliser of $|\Psi\rangle$ in the group $G_3.$ Except for a
set of states $|\Psi\rangle$ whose dimension is less than that of the
full space of states, 
\begin{equation}
 L(\Psi) = 0. 
\end{equation}
\end{thm}

\begin{proof}

Any state of three spin-$\tfrac{1}{2}$ particles is of the form:
\begin{equation*}
  |\Psi\rangle =
  \sum_{i,j,k}t_{ijk}|\psi_i\rangle|\psi_j\rangle|\psi_k\rangle 
\end{equation*}
where $i,j,k = 1$ or $2$ and $|\psi_1\rangle = |\up \rangle, \quad |\psi_2\rangle
= |\dn \rangle$.  A local transformation is of the form:
\begin{equation*}
  |\Psi\rangle \mapsto
  e^{i\varphi}\sum_{i,j,k,\ell,m,n}t_{ijk}u_{\ell i} v_{mj}w_{nk}
               |\psi_{\ell}\rangle |\psi_m\rangle |\psi_n\rangle 
\end{equation*}
for some $2 \times 2$ matrices $U,V,W \in SU(2)$ and some phase
$\varphi$.  Suppose $U,V,W$ are close to the identity:
\begin{equation}\label{epset}
    U = 1 + i\varepsilon A, \quad \quad 
    V = 1 + i\varepsilon B, \quad \quad
    W = 1 + i\varepsilon C 
\end{equation}
where $\varepsilon$ is infinitesimal and $A,B,C$ are hermitian and
traceless.  If $\theta = \varepsilon\varphi$ is also small we have, to
first order in $\varepsilon,$
\begin{align*}
  \delta|\Psi\rangle &= i \varepsilon\sum\left(\varphi t_{ijk} +
  a_{i\ell} t_{{\ell}jk} + b_{jm} t_{imk} + c_{kn} t_{ijn}\right)|\psi_i\rangle|\psi_j\rangle|\psi_k\rangle \\
  &= i \varepsilon\sum\left((\varphi \delta_{i\ell}  + a_{i\ell}) t_{{\ell}jk} + b_{jm} t_{imk} +
  c_{kn} t_{ijn}\right)|\psi_i\rangle|\psi_j\rangle|\psi_k\rangle 
\end{align*}
Hence if the local transformation
$\left(e^{i \theta},U,V,W\right)$ belongs to the stabiliser of
$|\Psi\rangle$, 
\begin{equation}\label{TL1}
  (\varphi \delta_{i\ell} + a_{i\ell}) t_{{\ell}jk} + b_{jm} t_{imk} + c_{kn} t_{ijn} = 0,
\end{equation}
using the summation convention on repeated indices.
Let $T_i$ be the matrix whose $(j,k)$th entry is $t_{ijk}$; then these
equations can be written in matrix form as 
\begin{equation}\label{TL2}
  (\varphi \delta_{i\ell} + a_{i\ell}) T_{\ell} + B T_i + T_i C\trans = 0.
\end{equation}
Separating these at their free indices, and performing the summation gives:
\begin{align*}
  B T_1 + T_1 C\trans + (\varphi + a_{11}) T_1 + a_{12} T_2 &=0 \\
  B T_2 + T_2 C\trans + a_{21} T_1 + (\varphi + a_{22}) T_2 &=0.
\end{align*}
Generically ({\bf{Gen 1}}), at least one of $T_1$ and $T_2$ is invertible
(say $T_2$); if so, 
\begin{align*}
 BT_1{T_2}^{-1} + T_1 C\trans {T_2}^{-1} + (\varphi +
a_{11})T_1{T_2}^{-1} + a_{12} &= 0 \\
T_1 {T_2}^{-1} B + T_1 C\trans {T_2}^{-1} + (\varphi + a_{22})T_1
{T_2}^{-1} + a_{21}\left(T_1 {T_2}^{-1} \right)^2 &= 0
\end{align*}
and so 
\begin{align}
  -T_1 C\trans {T_2}^{-1} &= B T_1 {T_2}^{-1} + (\varphi +              
       a_{11})T_1 {T_2}^{-1} + a_{12}\label{CTcond} \\
       &= T_1 {T_2}^{-1}B +(\varphi+a_{22})T_1{T_2}^{-1} +                 
       a_{21}\left( T_1 {T_2}^{-1} \right)^2\label{CTcond2}  
\end{align}
Let $X = T_1 {T_2}^{-1}$; then these equations give  
\begin{equation}\label{BcommX}
  \left[B,X\right] = -a_{12} \mathbf{1} - \left(a_{11}-a_{22}\right)X + a_{21}X^2
\end{equation}
Now we use
\begin{equation}\label{powertrace}
  \text{tr}\left(X^n\left[B,X\right]\right) = 
\text{tr}\left(X^nBX-X^{n+1}B\right) = 0
\end{equation}
to obtain
\begin{equation}\label{trace1}
  \text{tr}\left(a_{12}\mathbf{1} + \left(a_{11}-a_{22}\right)X - a_{21}X^2\right) = 0
\end{equation}
and
\begin{equation}\label{trace2}
  \text{tr}\left(a_{12}X + \left(a_{11}-a_{22}\right)X^2 - a_{21}X^3\right) = 0.
\end{equation}
Let $\lambda$ and $\mu$ be the eigenvalues of X.  Then we obtain 
\begin{align}
  2a_{12} + \left(a_{11}-a_{22}\right)\left(\lambda + \mu\right) -
  a_{21}\left(\lambda^2 + \mu^2\right) &= 0 \label{lambdatrace1} \\ 
  a_{12}\left(\lambda + \mu\right) +
  \left(a_{11}-a_{22}\right)\left(\lambda^2 + \mu^2\right) -
  a_{21}\left(\lambda^3+\mu^3\right) &=0 \label{lambdatrace2}
\end{align}
Generically ({\bf{Gen 2}}), $\lambda + \mu \neq 0$ and so solving for $a_{12}$
and $a_{21}$ in terms of $(a_{11}-a_{22})$ gives 
\begin{align}\label{a12eigs}
  &a_{12} = -\frac{\lambda \mu}{\lambda + \mu}\left(a_{11}-a_{22}\right),
  &a_{21} = \frac{1}{\lambda + \mu}\left(a_{11}-a_{22}\right), 
\end{align}
but generically ({\bf{Gen 3}}) this will not satisfy $a_{12} = \overline{a}_{21}$ unless
\begin{equation}\label{alphacond}
  a_{12} = a_{21} = (a_{11} - a_{22}) = 0.
\end{equation}
Hence $A = 0$, and the equations for $B$ and
$C$ become  
\begin{align}\label{BTTC1}
  \left(B - \varphi \mathbf{1}\right) T_1 + T_1 C\trans &= 0, \\
  B T_2 + T_2 \left(C\trans - \varphi \mathbf{1}\right) &= 0. 
\end{align}
The second of these equations gives $C$ as
\begin{equation}\label{BTTC2}
  C\trans = - {T_1}^{-1}\left(B- \varphi \mathbf{1} \right)T_1 = - {T_2}^{-1}B T_2 +
\varphi \mathbf{1}.
\end{equation}
Taking the trace of this equation, $\varphi = 0.$  Putting this into the first equation shows
that $B$ commutes with $T_1 {T_2}^{-1}$. Generically ({\bf{Gen 4}}), the only matrices which
commute with a $2 \times 2$ matrix $X$ are $\alpha 1 + \beta X$ for some scalars $\alpha$,
$\beta$; therefore
\begin{equation}\label{CayleyH}
  B = \alpha\mathbf{1} + \beta T_1{T_2}^{-1}.
\end{equation}
Generically ({\bf{Gen 5}}), this will not be hermitian unless $\beta=0$, and then tr$B =
0$ implies $\alpha = 0$. Thus $B=0$ and therefore
$C=0.$  Thus for generic values of $t_{ijk}$ the only solution of \eqref{TL2}
is 
\begin{equation}
 \varphi \mathbf{1} = A = B = C = 0,
\end{equation}
so the stability group is discrete.
\end{proof}

\paragraph{Remark:} 

It follows from this theorem that the generic orbit has the same
dimension as the group $G_3,$ namely $10.$  Since the space of
(non-normalised) state vectors has (real) dimension $16,$ the number of
independent invariants, which is the same as the dimension of the space
of orbits, is $6$ (including the norm). 

\section{Exceptional States: The significance of an enlarged stabiliser}

In this section we will prove that a three-qubit state which is
exceptional in the sense of Theorem 1 has a stationary value of some
fundamental invariant.  Since any measure of entanglement must be such
an invariant, this indicates that these mathematically exceptional
states are likely to have a special physical significance. 

By a ``local invariant'' we mean a real-valued function of the state
vector which is invariant under local unitary transformations, and is
therefore constant on each orbit.  It is convenient to concentrate on
polynomial functions, which can be regarded as coordinates on the space
of entanglement types; more general invariants (e.g. entropy of
entanglement) can be constructed from these. Since the generic orbit in
the state space $\mathcal{H}$ has dimension dim$G_3,$ the number of parameters
needed to specify such an orbit is dim$\mathcal{H} -$ dim$G_3 = 6.$ 
Such parameters, being constant on orbits, are invariants. 

The space of orbits is not necessarily flat, and it may not be possible
to parametrise it globally with a single set of six invariants (see
\cite{Grassl}): geometrically, the space of orbits is a manifold which
may have several different coordinate patches; algebraically, the
algebra of invariants is not a polynomial algebra but is generated by
more than six invariants which are subject to some relations.  However,
we can choose a neighbourhood of a state so that the algebra of
invariant functions on that neighbourhood has six independent generators. 

\begin{thm}\label{elastic}
Let $\mathcal{H}$ be the space of $3$-qubit pure states, and let
$G_3$ be the group of local unitary transformations of
$\mathcal{H}.$   Let $I_1,...,I_6$ be a set of
$6$ polynomial invariants which generate the algebra of local invariants
in a neighbourhood of a state $|\psi_0\rangle.$  If the stabiliser of
$|\psi_0\rangle$ in $G_3$ has non-zero dimension, there is a
linear combination of $I_1,...,I_6$ which has a stationary value at
$|\psi_0\rangle.$   
\end{thm}

\begin{proof}
Let $x_1,...,x_{16}$ be real coordinates on $\mathcal{H}.$ Suppose the
Jacobian matrix
\begin{equation}
 J= \left(\frac{\partial I_i}{\partial x_j}\right)
\end{equation}
has maximal rank 6 at $|\psi_0\rangle.$  Since the $I_i$ are
polynomials, the $6 \times 6$ minors of $J$ are continuous
functions, so if one of them is non-zero at $|\psi_0\rangle$ it is
non-zero in a neighbourhood of $|\psi_0\rangle.$  Hence, by the implicit
function theorem, the equations 
\begin{equation}\label{level} 
 I_i \left(|\psi \rangle \right) = I_i \left(|\psi_0 \rangle \right)
\end{equation}
define a smooth manifold in $\mathcal{H}$ of dimension dim$\mathcal{H} -
6.$  These are the equations of a level set of the polynomial invariants
of $G_3.$ Since $G_3$ is compact, its invariants separate
the orbits \cite{maxref} and so \eqref{level} is the equation of the orbit
of $|\psi_0\rangle,$ which therefore has the same dimension as
$G_3.$  It follows that the stabiliser of $|\psi_0\rangle$ is
discrete.  

Hence if the stabiliser of $|\psi_0\rangle$ is {\emph{not}} discrete,
then the matrix $J$ has rank less than $6$ and therefore
there exist scalars $(\lambda_1,...,\lambda_6)$ such that
\begin{equation}\label{lambdacond}
 \sum_i^6 \lambda_i \frac{\partial I_i}{\partial x_j} \left(|\psi_0\rangle
\right) = 0,
\end{equation}
i.e., the linear combination
\begin{equation}
 \sum \lambda_i I_i
\end{equation}
has a stationary value at $|\psi_0\rangle.$
\end{proof}

Note that this theorem does not guarantee that all stationary subspaces
of any invariant will be associated with enlarged stabilisers.  However,
it does indicate that states with enlarged stabiliser dimensions are
likely to have special physical significance.  

\section{The classification of non-generic states}

\subsection{Setting up the problem.}

We will look for the stabilising subgroup of the group $G_3 = U(1) \times
SU(2)^3$ of local transformations, i.e. the group of $(e^{i\varphi},U,V,W)$
where $U,V,W$ are all elements of $SU(2)$ and $e^{i\varphi}$ is an
overall phase.  We will start with the three-index
tensor equation for the local transformations:
\begin{equation}\label{tensor1} 
  {t'}_{ijk} = \sum e^{i\varphi} u_{il} v_{jm} w_{kn} t_{lmn}
\end{equation}
where the $t$'s are the coefficients of the state vector and the
$u_{il}$'s  are the matrix elements of $U \in SU(2)$ etc. Using the
${(T_i)}_{jk}$ notation introduced in Theorem 1, and partitioning the
equation at the index $i$:
\begin{equation}\label{transf1}
 {T'}_1 = e^{i\varphi}V\left[ u_{11} T_1 + u_{12} T_2 \right] W\trans
\end{equation}
\begin{equation}\label{transf2}
 {T'}_2 = e^{i\varphi} V \left[ u_{21} T_1 + u_{22} T_2 \right] W\trans   
\end{equation}  
where $u_{22} = \overline{u_{11}}$ and $u_{21} = -\overline{u_{12}}$ and
$|u_{11}|^2 + |u_{12}|^2 = 1$.  The stabiliser is the set of  
$\left(e^{i\varphi} , U , V , W \right)$ such that
${T'}_1 = T_1$ and ${T'}_2 = T_2.$  

In examining the non-generic states, not covered by Theorem \ref{tony's},
whose stabilisers have potentially non-zero dimension, we will sometimes
find it convenient to abandon the infinitesimal approach of Theorem 1 and
determine all finite elements of the stabiliser groups.

\subsection{The ``bystander'' rule.}
  We will now examine the apparently trivial case when either $T_i$ (say
$T_1$) is the zero matrix.   
In this instance it is possible to choose bases of the two one-particle
spaces such that $T_1$ and $T_2$ become diagonal.  We need
therefore only consider the case  
\begin{equation*}
  T_2 =
  \begin{pmatrix} \alpha & 0 \\
                  0     & \beta
  \end{pmatrix}
\end{equation*}
where $\beta$ may or may not be zero.  Then the first stabiliser equation
\eqref{transf1} becomes 
\begin{equation*}
  0 = e^{i\varphi} V \left( u_{11} 0 + u_{12} T_2 \right) W\trans
\end{equation*}  
therefore $u_{12} = u_{21} = 0$, and the other equation becomes 
\begin{equation}\label{bystander}
  T_2 = e^{i\varphi} V \left( e^{\pm i\theta} T_2 \right) W\trans  
\end{equation}
where $e^{i\theta} = u_{22}.$  This can be seen
to be the 2-particle stabiliser equation, but with an additional
external phase factor -- which for the sake of transparency later we
will not absorb into $\varphi$. 
  The fact that one of the $T_i$-matrices is the zero matrix means that
states of this type are factorisable.  The particle(s) whose kets can be
factored out in this way do not participate in the entanglement (if any) of
the other particles and so we'll call these `bystander' particles, and
states in which not all the particles participate in the entanglement
`bystander' states.

If $T_2$ is singular, we have the equation
\begin{equation*}
  \begin{pmatrix} \alpha & 0 \\
                   0     & 0     
  \end{pmatrix}
  = e^{i\varphi} e^{i\theta} V   
  \begin{pmatrix} \alpha & 0 \\
                   0     & 0 
  \end{pmatrix}
  W\trans 
\end{equation*}
which, by the two-particle result reduces to $V=e^{i\gamma
\sigma_{3}},W=e^{i\eta \sigma_{3}}$ with
\begin{equation}\label{byst1}
  \alpha = e^{i\varphi} e^{i\theta} e^{i\gamma} \alpha e^{i\eta}
\end{equation}
giving us the condition $$\varphi + \theta + \gamma + \eta = 2n \pi$$  
i.e., three degrees of freedom. 

If $T_2$ is non-singular, use Section 2 to look up the appropriate
2-particle stabiliser.  This comes down to whether or not $|\alpha| =
|\beta|$.  If $|\alpha| \neq |\beta|$, the equation becomes
\begin{align*}
  \alpha &= e^{i\varphi} e^{i\theta} e^{i\gamma} \alpha e^{i\eta} \\
  \beta &= e^{i\varphi} e^{i\theta} (-e^{i\gamma}) \beta (-e^{i\eta})  
\end{align*}
which both reduce to 
\begin{equation}\label{bystcond}
  \varphi + \theta + \gamma + \eta = 2n \pi,
\end{equation}
which makes three degrees of freedom. 

If $|\alpha| = |\beta|$ , take $\alpha = \beta \in \mathbb{R}$.
Then we have 
\begin{equation}\label{bystmax}
  \alpha \mathbf{1} = e^{i\varphi} e^{i\theta} U \alpha \mathbf{1} U^\dagger  
\end{equation}
so $$\varphi + \theta = 2n \pi$$ and one element of $SU(2)$
giving us four degrees of freedom. 

Thus (in the three spin-$\tfrac{1}{2}$ case) factorisable states
reproduce the stabilising group structure of the
fewer-particle states that their sub-systems resemble.

\subsection{Exchanging the particle labels.}

Recall that in Theorem 1 we chose particle 1, with corresponding index
$i$, as the `partitioning index' which splits the original, 3-index state
vector `tensor' problem into the more manageable form of a pair of coupled
matrix equations.  
\begin{equation}\label{partition}
 \mathcal{P}_i : t_{ijk} \to (T_i)_{jk} 
\end{equation}
This choice of particle 1 was entirely arbitrary: we
could just as easily have chosen either of the indices $j$ or $k$.
Changing the partition index is sometimes
useful. The effect of changing the particle labels (repartitioning) on the
stabiliser is simply to permute $U,V,W$ as each particle's associated $SU(2)$
copy just follows its associated index.  

In group theoretical terms, the operations of permuting the particles
are unitary operations on three-particle states which, though not
elements of the group of {\emph{local}} unitary transformations, do
belong to the normaliser of this subgroup in the group of all unitary
transformations.  States related by elements of the normaliser will have
isomorphic stabilisers in the group of local unitary transformations.

\subsection{Change of basis}

We are, of course, always free to change the basis that we use to
describe states of any of the three particles.  (This amounts to
applying a local unitary transformation in the passive interpretation.)
If the change of basis is described by the $2\times2$ matrix $P$ for
particle 1, $Q$ for particle 2 and $R$ for particle 3, then the effect
on the matrices $U,V,W$ is
\begin{align}
 &U \to PUP^{-1}, 
 &V \to QVQ^{-1}, \quad \quad \quad
 &W \to RWR^{-1}.
\end{align}
The effect on the matrices $T_1,T_2$ is the same as in \eqref{transf1},
\eqref{transf2} with $(P,Q,R)$ replacing $(U,V,W).$  In other words, the
group element $(e^{i\varphi},U,V,W)$ is conjugated by the group element
corresponding to $(P,Q,R)$ (namely $(e^{i\theta},P',Q',R')$ where
$e^{i\theta}=(\det P \cdot \det Q \cdot \det R)^{\frac{1}{2}}$ and
$P'=(\det P)^{-\frac{1}{2}}P,$ etc.)

If we regard $P,Q,R$ as active transformations, taking the state
$(T_1,T_2)$ to a different state on the same orbit, then this is the
basis of our earlier remark that all the points on a given orbit have
conjugate stabilisers.

\subsection{Type 1 Non-Generic States: Both $T$-matrices singular}

In Theorem \ref{tony's} the first step in the argument that is only
generically true ({\bf{Gen 1}}) needs at least one $T_i$ to be
invertible for the argument to be valid.  If both $T_i$'s are singular,
we can choose our coordinates to put one $T_i$, $T_1$ say, into diagonal
form by an appropriate local transformation. Then $T_1$ and $T_2$ will
be of the form 
\begin{align}\label{semistand}
     &T_1 = 
     \begin{pmatrix} p & 0 \\
                     0 & 0
     \end{pmatrix}, 
     &T_2 =
     \begin{pmatrix} ac & ad \\
                     bc & bd 
     \end{pmatrix}, 
\end{align}
where the singular value $p$ is real and positive (the case where $p=0$
has already been dealt with in subsection 5.2.)  
The stabiliser equations, obtained from \eqref{transf1} and
\eqref{transf2} by imposing the conditions that ${T'}_1=T_1$ and ${T'}_2
= T_2$, are:
\begin{equation}\label{stab1}
 T_1 = e^{i\varphi} V \left[u_{11} T_1 + u_{12} T_2 \right] W\trans 
\end{equation}
\begin{equation}\label{stab2}
 T_2 = e^{i\varphi} V \left[\overline{u}_{11} T_2 - \overline{u}_{12} T_1 \right] W\trans.
\end{equation}

From \eqref{stab1} and \eqref{stab2} it can be seen that a necessary
condition for an enlarged stabiliser to occur is that $u_{11}T_1 +
u_{12}T_2$ and $-\overline{u}_{12}T_1 + \overline{u}_{11}T_2$ must have
the same singular values as $T_1$ and $T_2$ respectively.  In
particular, they must have the same determinant, namely zero.  Taking the
determinant of $u_{11}T_1 + u_{12}T_2$,
\begin{equation}\label{detcond}
  u_{11} u_{12} pbd = 0.
\end{equation}

We will write
\begin{align}\label{genunks}
 &V = 
    \begin{pmatrix} g & h \\
                    -\overline{h} & \overline{g} 
    \end{pmatrix}, \quad
 &W = 
    \begin{pmatrix} r & s \\
                    -\overline{s} & \overline{r} 
    \end{pmatrix}
\end{align} 

\subsubsection{Case 1: $a,b,c,d$ all nonzero (Semigeneric states)}

Suppose $a,b,c,d$ are all non-zero. We will call this form
``Semigeneric'', as it is the generic form for a singular matrix for
$T_2$. Equation \eqref{detcond} shows that either
$u_{11}=0$ or $u_{12}=0.$  If $u_{12}=0,$ write $u_{11}=e^{i\theta};$
then \eqref{stab1} becomes
\begin{equation}
 \begin{pmatrix} p & 0 \\
                 0 & 0 
 \end{pmatrix}
 = 
 pe^{i(\varphi+ \theta)}
 \begin{pmatrix} gr & -g\overline{s} \\
                 -\overline{h}r & \overline{h}\overline{s} 
 \end{pmatrix}.
\end{equation}
Hence $h=s=0, \quad g=e^{i\alpha}, \quad r=e^{i\beta},$ with 
\begin{equation}\label{semig1}
 \varphi + \theta + \alpha + \beta = 0 \text{ or } 2\pi. 
\end{equation}

Now \eqref{stab2} gives
\begin{equation}
 \begin{pmatrix} ac & ad \\
                 bc & bd 
 \end{pmatrix}
 = e^{i(\varphi - \theta)}
 \begin{pmatrix} e^{i(\alpha + \beta)}ac & e^{i(\alpha - \beta)}ad \\
                 e^{i(-\alpha + \beta)}bc & e^{-i(\alpha + \beta)}bd
 \end{pmatrix},     
\end{equation}
so
\begin{equation}\label{semig2}
 \alpha + \beta = \alpha - \beta = -\alpha + \beta = -\alpha -\beta =
\theta - \varphi \quad (\text{mod }2\pi).
\end{equation}
From this, together with \eqref{semig1}, it follows that each of the
angles $\varphi,\theta,\alpha,\beta$ is equal to $0$ or $\pi$ and
therefore the stabiliser is discrete.  

We will write the stabiliser as $\mathcal{S}=\mathcal{S}_1 \cup
\mathcal{S}_2,$ where $\mathcal{S}_1$ is the subset with $u_{12}=0$ and
$\mathcal{S}_2$ is the subset with $u_{11}=0.$  Then $\mathcal{S}_1$ is
a subgroup.  The product of any two elements of $\mathcal{S}_2$ belongs
to $\mathcal{S}_1,$ so $\mathcal{S}_2$ is a single coset of
$\mathcal{S}_1$ (unless it is empty) and therefore contains the same
number of elements as $\mathcal{S}_1,$ and is therefore also discrete.

\paragraph{Case 2: $a=0$ or $b=0$, $bd \neq 0$ (Slice states)}

If either $a$ or $c=0$ and the other three of $a,b,c,d$ are non-zero, then the
state is either
\begin{align}
 &p|\up \up \up \rangle + bc|\dn \dn
\up \rangle + bd|\dn \dn \dn \rangle \label{slice1} \\
 \text{or } \quad &p|\up \up \up \rangle + ad|\dn
\up \dn \rangle + bd|\dn \dn \dn
\rangle \notag
\end{align}
which are equivalent to each other under exchange of particles 2 and 3.
(The third similar state,
\begin{equation}
 p|\up \up \up \rangle + q|\up \dn \dn
\rangle + r|\dn \dn \dn \rangle
\end{equation}
can be obtained by a permutation of the particle labels.)  For the state
\eqref{slice1} the equations for $u_{12}=0$ give the one-dimensional set
of stabiliser elements
\begin{equation}\label{straightslice}
 (e^{i\varphi},U,V,W) = \left( \varepsilon_1 \mathbf{1},e^{i\theta
\sigma_3},\varepsilon_2 e^{-i\theta \sigma_3}, \varepsilon_1
\varepsilon_2 \mathbf{1} \right) \quad \text{ where } \quad \varepsilon_1,
\varepsilon_2 = \pm 1.
\end{equation}

The equations for $u_{11}=0$ require $T_1$ and $T_2$ to have the same
singular values, the condition for which is 
\begin{equation}\label{sliceflipcond}
 p^2 = |b|^2 \left(|c|^2 + |d|^2 \right).
\end{equation}
If this is satisfied, the stabiliser equations are 
\begin{equation}
 \begin{pmatrix} 0 & 0 \\
                 bc & bd 
 \end{pmatrix}
 = p e^{-i(\varphi + \theta)}
 \begin{pmatrix} \overline{g}\overline{r} & \overline{g}\overline{s} \\
                 \overline{h}\overline{r} & \overline{h}\overline{s}
 \end{pmatrix} 
 = -p e^{i(\varphi -\theta)}
 \begin{pmatrix} gr & -g\overline{s} \\
                 -\overline{h}r & \overline{h}\overline{s}
 \end{pmatrix}
\end{equation}
These give the stabiliser elements with $u_{11}=0$ as
\begin{multline}\label{sliceflip}
 \left(e^{i\varphi},U,V,W \right) = \\
 \left( \varepsilon_1 i,
 \begin{pmatrix} 0 & e^{i\theta} \\
                 -e^{-i\theta} & 0 
 \end{pmatrix}, 
 \varepsilon_2
 \begin{pmatrix} 0 & e^{-i(\theta + \chi)} \\
                 -e^{i(\theta + \chi)} & 0 
 \end{pmatrix},
 \varepsilon_1 \varepsilon_2
 \begin{pmatrix} -i \frac{|bc|}{p} & -i\frac{\overline{b}\overline{d}}{p}e^{i\chi} \\
                 -i\frac{bd}{p}e^{-i\chi} & i\frac{|bc|}{p}
 \end{pmatrix}
 \right) 
\end{multline}
where $\chi = $arg$(bc)$ and $\theta$ can take any value between $0$ and
$2\pi.$  

Thus the slice states have a one-dimensional stabiliser consisting of
the four circles \eqref{straightslice} unless \eqref{sliceflipcond} is
satisfied, when the stabiliser is doubled and also contains the four
circles \eqref{sliceflip}.  We call this set of states a ``slice ridge''.

\paragraph{Case 3: $a=c=0, \quad bd \neq 0$ (The GHZ states)}

If $a=c=0$, but $bd \neq 0$ the state is the GHZ state. 
\begin{equation}
 p|\up \up \up \rangle + q|\dn \dn
\dn \rangle 
\end{equation}
with $p$ and $q=bd$ both non-zero.  We may assume that they are both
real and positive.  The singular value condition tells
us that unless $|q|=p$ the only solutions to the stabiliser equations
will have $u_{12}=0,$ giving the two-dimensional stabiliser
\begin{equation*}
  \left(e^{i\varphi},U,V,W \right)=
  \left(\pm \mathbf{1}, e^{i\theta \sigma_3}, e^{i\alpha \sigma_3},           
   e^{i\beta \sigma_3} \right) 
\end{equation*}
with the condition that $\theta + \alpha + \beta = 0$ or $\pi.$   

If $|q|=p$, the stabiliser is doubled, and also contains the elements
\begin{equation*}
  \left(e^{i\varphi},U,V,W \right)=  
  \left(\pm i, 
  \begin{pmatrix} 0 & e^{i\theta} \\
                  -e^{-i\theta} & 0 
  \end{pmatrix},
  \begin{pmatrix} 0 & e^{i\alpha} \\
                  -e^{-i\alpha} & 0 
  \end{pmatrix},
  \begin{pmatrix} 0 & e^{i\beta} \\
                  -e^{-i\beta} & 0 
  \end{pmatrix}
  \right)             
\end{equation*}
with the condition that 
\begin{equation}
\theta + \alpha + \beta = 0 \text{ or } \pi
\end{equation}
This is the original GHZ state, which can be regarded as
a three-particle analogue of the maximally entangled (``singlet'')
two-particle state.  We note that although the GHZ state has an enlarged
stabiliser when its coefficients are equal in magnitude, the enlargement
does not consist of an increase in dimension as in the two-particle
case.

\subsubsection{Case 4: $b=0$ or $d=0$ (Bystander states)}   

If $b$ or $d$ or both are zero, the determinant equation \eqref{detcond}
no longer implies that $U$ must be either diagonal or anti-diagonal. 
However, in all of these cases the state factorises and one of the
particles is a bystander.  We will just look at the $b=0$ case, as $d=0$
can be obtained by the appropriate transpositions, and go back to the
``both'' case after that.  We have the state vector:
\begin{align*}
&T_1 =
\begin{pmatrix} p & 0 \\
                0 & 0     
\end{pmatrix}, 
&T_2 = 
\begin{pmatrix} ac & ad \\
                0  & 0 
\end{pmatrix} 
\end{align*}
i.e.,
\begin{equation*}
 |\up_2 \rangle \left( p|\up_1 \up_3 \rangle + ac|\dn_1 \up_3 \rangle
+ ad|\dn_1 \dn_3 \rangle \right)
\end{equation*}
which is a state in which particle 2 is a bystander, and therefore has
been dealt with in section 5.2 above.

\subsubsection{Case 5: $b=d=0$ (completely factorised states)}

In this case, 
\begin{align*}
&T_1 =
\begin{pmatrix} p & 0 \\
                0 & 0
\end{pmatrix},   
&T_2 = 
\begin{pmatrix} ac & 0 \\
                0  & 0 
\end{pmatrix}  
\end{align*}
so the state vector is:
\begin{equation}
 \left(p|\up_1\rangle + ac|\dn_1 \rangle\right)|\up_2
\up_3 \rangle
\end{equation}
which is the totally factorised state, and has already been considered
as the $T_2$ singular bystander case.

\subsection{Non-generic Type 2: tr$(T_1{T_2}^{-1})=0$.}

Let us now consider
what might happen if the assumption ({\bf{Gen 2}}) fails. 
If $\lambda + \mu = 0,$ equations \eqref{trace1} and \eqref{trace2} become
\begin{align}\label{ng2cond}
 a_{12} &= \lambda^2 a_{21}, \\
 2\lambda^2(a_{11} - a_{22}) &= 0.
\end{align}
We can still deduce that $A=0$ (since $a_{12}=\overline{a}_{21}$ and
$a_{11}+a_{22}=0$) unless $|\lambda|=1$ or $\lambda = \mu = 0.$

\subsubsection{Case 1: $|\lambda|=1.$}

Since $a_{12}=\overline{a}_{21},$ equation \eqref{ng2cond} gives $a_{12}
= \alpha \lambda$ where $\alpha$ is real.  The right hand side of
\eqref{BcommX} becomes
\begin{equation}
 \alpha \overline{\lambda}\left(X^2 -\lambda^2 \mathbf{1} \right)
\end{equation}
by the Cayley-Hamilton theorem.  Thus it is still true that $B$ must
commute with $X.$  We can change basis for particle $2$ (multiplying
$T_1$ and $T_2$ on the left by a unitary matrix $P$) so that $X$ takes
the form
\begin{equation}\label{Xinit}
 X = T_1 {T_2}^{-1} = 
 \begin{pmatrix} \lambda & \omega \\
                 0 & -\lambda
 \end{pmatrix}.  
\end{equation}
Since $X$ is not a multiple of the identity, the requirement that $B$
should commute with $X$ gives
\begin{equation}\label{Buv}
 B = u\mathbf{1} + v X
\end{equation}
for some scalars $u,v;$ but $B$ is traceless, so $u=0.$

Suppose $\omega \neq 0.$  Since $B$ is hermitian, $v=0;$ thus $B=0.$ 
Now equation \eqref{CTcond2} gives
\begin{equation}
 C\trans = -\varphi\mathbf{1}-\alpha\overline{\lambda}{T_2}^{-1}T_1.  
\end{equation}
Hence
\begin{align}
 \varphi &= -\frac{1}{2} \text{tr}\left[C\trans + \alpha
\overline{\lambda} {T_2}^{-1} T_1 \right] \\
         &= -\frac{1}{2} \text{tr} \left[\alpha \overline{\lambda} 
                                    T_1 {T_2}^{-1} \right] = 0.
\end{align}
Now we can change basis for particle $3$ (multiplying $T_1$ and $T_2$ on
the right by a unitary matrix) so that $T_2$ takes the form 
\begin{equation}\label{vorpal}
 T_2 =
 \begin{pmatrix} a & b \\ 
                 0 & 1
 \end{pmatrix}
\end{equation}
with $a \neq 0$ since $T_2$ is invertible.  Then 
\begin{align}
 C\trans &= -\alpha \overline{\lambda} {T_2}^{-1} X T_2 \\
         &= -\alpha
 \begin{pmatrix} 1 & a^{-1}(\overline{\lambda}\omega + 2b) \\
                 0 & -1 
 \end{pmatrix}.
\end{align}
Since $C$ is hermitian, a non-discrete stabiliser can only occur if 
\begin{equation}
 \overline{\lambda} = - \frac{2b}{\omega}.
\end{equation}
Then the state is
\begin{align}\notag
 |\Psi \rangle &= \lambda |\up \rangle \left(a|\up\rangle |\up \rangle
            - b|\up\rangle |\dn\rangle - |\dn \rangle
              |\dn \rangle \right) \\
       &+ |\dn\rangle\left(a|\up\rangle|\up\rangle + b
          |\up\rangle|\dn \rangle + |\dn\rangle |\dn
          \rangle \right) \\
       &= a|\up'\rangle|\up\rangle\up\rangle +
          b|\dn' \rangle |\up \rangle |\dn \rangle 
          - \lambda |\dn'\rangle |\dn\rangle |\dn
            \rangle \label{prime1}
\end{align}
where 
\begin{align*}
 |\up' \rangle &= \frac{1}{\sqrt{2}}\left(\lambda|\up\rangle 
                                               + |\dn\rangle\right) , \\
 |\dn' \rangle &= \frac{1}{\sqrt{2}}\left(|\up\rangle       
                              - \overline{\lambda}|\dn\rangle\right). 
\end{align*}
This is one of the slice states considered in Section 5.5.

If $\omega = 0,$ equations \eqref{Xinit} and \eqref{vorpal} immediately
give
\begin{align}\notag
 |\Psi \rangle &= \lambda |\up \rangle \left(a|\up\rangle |\up \rangle
            + b|\up\rangle |\dn\rangle - |\dn \rangle
              |\dn \rangle \right) \\
       &+ |\dn\rangle\left(a|\up\rangle|\up\rangle + b
          |\up\rangle|\dn \rangle + |\dn\rangle |\dn
          \rangle \right) \\
       &= a|\up'\rangle|\up\rangle\up\rangle +
          b|\dn' \rangle |\up \rangle |\dn \rangle 
          - \lambda |\dn'\rangle |\dn\rangle |\dn \rangle \label{prime2}
\end{align}
which is again a slice state.

\subsubsection{Case 2: $\lambda = \mu = 0$.}

The only remaining possibility is that $X=T_1{T_2}^{-1}$
is unitarily equivalent to  
\begin{equation}
 \begin{pmatrix} 0 & \omega \\
                 0 & 0 
 \end{pmatrix}.
\end{equation}
In this case \eqref{trace1} and \eqref{trace2} give only
$a_{12}=a_{21}=0$ and \eqref{BcommX} becomes
\begin{equation}
 BX - XB = 2rX
\end{equation}
where $r=a_{11}=-a_{22}.$  With $X={\textstyle{\begin{pmatrix} 0 & \omega \\ 0 & 0
\end{pmatrix}}},$ it follows that 
\begin{equation}
 B = 
 \begin{pmatrix} r & 0 \\
                 0 & -r
 \end{pmatrix},
\end{equation} 
i.e., $B=A.$  Now we return to equations \eqref{TL2} of theorem
\ref{tony's}:
\begin{equation*}
 (\varphi \delta_{i\ell} + a_{i\ell}) T_{\ell} + B T_i + T_i C\trans = 0.
\end{equation*}
Writing 
\begin{equation*}
 T_2 = 
 \begin{pmatrix} a & b \\
                 c & d 
 \end{pmatrix}
\end{equation*}
so that 
\begin{equation*}
 T_1 =
 \begin{pmatrix} \omega c & \omega d \\
                 0 & 0 
 \end{pmatrix}
\end{equation*}
and 
\begin{equation*}
 C\trans =
 \begin{pmatrix} s & y \\
                 \overline{y} & -s 
 \end{pmatrix}
\end{equation*}
these become:
\begin{equation*}
  \begin{pmatrix} r & 0 \\
                  0 & -r
  \end{pmatrix}
  \begin{pmatrix} \omega c & \omega d \\
                  0        & 0
  \end{pmatrix}
  +
  \begin{pmatrix} \omega c & \omega d \\
                  0        & 0
  \end{pmatrix}
  \begin{pmatrix} x       & y \\
                  \overline{y} & -x 
  \end{pmatrix} 
  = -(\varphi + a_{11})
  \begin{pmatrix} \omega c & \omega d \\
                  0        & 0
  \end{pmatrix}              
\end{equation*}
and 
\begin{equation*}
  \begin{pmatrix} r & 0 \\ 
                  0 & -r
  \end{pmatrix}
  \begin{pmatrix} a & b \\
                  c & d
  \end{pmatrix}
  +
  \begin{pmatrix} a & b \\
                  c & d 
  \end{pmatrix}
  \begin{pmatrix} x       & y \\
                  \overline{y} & -x
  \end{pmatrix}
  = -(\varphi - a_{11}) 
  \begin{pmatrix} a & b \\
                  c & d 
  \end{pmatrix}.
\end{equation*}  
These give us the following constraints:
\begin{align*}
  c(s+r) + d \overline{y} &= -c(\varphi + r) \\
  d(r-s) + cy &= -c(\varphi + r) \\
  a(r+s) + b \overline{y} &= -a(\varphi - r) \\
  b(r-s) + ay &= -b(\varphi - r) \\
  c(s-r) + d \overline{y} &= - c(\varphi - r) \\
  -d(r+s) + cy &= -d(\varphi - r). 
\end{align*}
which produce just four independent equations:
\begin{align}\label{singset}
  \frac{a}{2}(\varphi + r) - \frac{3a}{2}(\varphi - r) - as - b\overline{y} &= 0 \\ 
  \frac{b}{2}(\varphi + r) - \frac{3b}{2}(\varphi - r) + bs - ay &= 0 \\
  \frac{c}{2}(\varphi + r) + \frac{c}{2}(\varphi - r) + cs + d\overline{y} &= 0 \\
  \frac{d}{2}(\varphi + r) + \frac{d}{2}(\varphi - r) - ds + cy &= 0.
\end{align}
For a non-zero solution $(\varphi,r,s,y)$ with $\varphi,r,s$ real, the
matrix  
\begin{equation}\label{matrix0}
  \begin{pmatrix} \overline{a} & -3\overline{a} & -2\overline{a}  & -2\overline{b} \\
                  b       & -3b       & 2b         & -2a       \\            
                  \overline{c} & \overline{c}   & 2\overline{c}   & 2\overline{d}  \\   
                  d       & d         & -2d        & 2c
  \end{pmatrix}
\end{equation}
must have determinant zero.  This gives us that
\begin{equation}\label{det0}
  \text{det}(T_2) \overline{a}\overline{c} + \text{det}(\overline{T_2}) bd = 0.   
\end{equation}
Since $T_2$ is non-singular by assumption, this allows us only three
possible solutions:
\begin{align}\label{possoln}  
  a &= d = 0 \\
  c &= b = 0 \\
  |ac| &= |bd|, \quad \text{all non-zero}.
\end{align}

If $a=d=0$ we have $b\overline{y}=0$ therefore $y=0$.  Then
\begin{align*}
  (r-s) &= -(\varphi - r) \\
  (s-r) &= -(\varphi - r)
\end{align*}
which gives us that $(\varphi - r) = 0$ and also that $s = r.$  This
solution has one degree of freedom, which we'll call $\varphi.$  The
state is:
\begin{align}\label{beechnut}
  &T_1 = 
  \begin{pmatrix} \omega c & 0 \\
                  0        & 0 
  \end{pmatrix},  
  &T_2 = 
  \begin{pmatrix} 0 & b \\  
                  c & 0 
  \end{pmatrix}  
\end{align}  
and the stabiliser for states of this type is,
\begin{equation}\label{beechstab}
  \left(e^{i\varphi},U,V,W \right) =
  \left(e^{i\varphi}, e^{i\varphi \sigma_3}, e^{i\varphi \sigma_3},
                      e^{-i\varphi\sigma_3} \right)
\end{equation}

If $b=c=0$ we have a state vector that looks like this:
\begin{align*}
   &T_1 = 
   \begin{pmatrix} 0 & \omega d \\
                   0 & 0 
   \end{pmatrix},
   &T_2 = 
   \begin{pmatrix} a & 0 \\
                   0 & d
   \end{pmatrix}
\end{align*}
which is just a reflection of the state vector in the previous case in
the vertical midlines, and so can be mapped into it by a change of basis,
as can its siblings obtained by permuting the particle labels.  The
stabiliser for these is thus:
\begin{equation*}
 \left(e^{i\varphi},U,V,W \right) =
  \left(e^{-i\varphi}, e^{-i\varphi \sigma_3}, e^{-i\varphi \sigma_3},
                      e^{i\varphi\sigma_3} \right)
\end{equation*}
so relabelling the spin coordinate just relabels the stabiliser variable,
as expected.  We nickname these states ``Beechnut'' states, because when
the three one-particle von Neumann entropies for this subspace are
plotted, we think it looks like a beech nut.

This leaves us with the ``non-zero'' solution. It can be seen that
$(\varphi + r) c = d \overline{y}$ and hence that $2rc =
-2rc$ which means that $r=0$ since we've assumed that $c \neq 0.$  Hence
$r=s=0$ and $b \overline{y} = \varphi a.$  So we have 
\begin{align*}
  b \overline{y} &= \varphi a \\ 
  \overline{y} &= \varphi \frac{a}{b} \\
  \overline{y} &=  \varphi \frac{c}{d}
\end{align*}
and so $$\frac{a}{b} = \frac{c}{d}.$$  Therefore $$ad=bc$$ and the
determinant of $T_2$ is zero after all: this case is Type 1
Non-generic, and is in fact a bystander case.

\subsection{Non-generic Type $3$}

In this next stage of the calculation, we will assume that both $T_1$ and
$T_2$ are non-singular, and move on to consider the failure of the
assumption ({\bf{Gen 3}}). 
In Theorem $1$ we obtained the equations \eqref{a12eigs}
\begin{align}
  &a_{12} = -\frac{\lambda \mu}{\lambda + \mu}(a_{11}- a_{22}) 
  &a_{21} = \frac{1}{\lambda + \mu}(a_{11} - a_{22}) 
\end{align}
where $\lambda, \mu$ are the eigenvalues of the matrix $X = T_1{T_2}^{-1}$.
But generically, this will not satisfy $a_{12}=\overline{a}_{21}$ unless
\begin{equation*}
  a_{12} = a_{21} = a_{11}-a_{22} = 0,
\end{equation*}   
so that $A=0.$  We will now examine values of $\lambda$ and $\mu$ that
allow $A$ to be non-zero.

Since $A$ is hermitian and traceless, $a_{11}=-a_{22}$ is real.  So
$a_{12}=\overline{a_{21}}$ requires
\begin{equation*}
 -\frac{\lambda \mu}{\lambda + \mu} = \frac{1}{\bar{\lambda} +
\overline{\mu}} 
\end{equation*}
i.e.,
\begin{equation*}
 -|\lambda|^2\mu - \lambda|\mu|^2 = \lambda + \mu.
\end{equation*}
Now we know that $|\lambda\mu|=1$ from these same equations. 
Substituting for $|\mu|^2$ gives
\begin{equation}
 \left(|\lambda|^2+1\right)\left(\mu |\lambda|^2 + \lambda \right) = 0
\end{equation} 
Hence
\begin{equation}
 \lambda \left(\overline{\lambda}\mu + 1 \right) = 0
\end{equation}
and so the eigenvalues of $T_1{T_2}^{-1}$ must be of opposite phase, namely:
\begin{align}
 &\lambda, \quad &-\frac{1}{\overline{\lambda}}. 
\end{align}

Writing $a_{11}=\alpha=-a_{22},$ we now have
\begin{equation}
 A = \alpha 
 \begin{pmatrix} 1 & \frac{2 \lambda}{|\lambda|^2 - 1} \\
                 \frac{2 \overline{\lambda}}{|\lambda|^2 - 1} & -1
 \end{pmatrix}.
\end{equation}
The right-hand side of \eqref{BcommX} becomes
\begin{equation}
 \frac{2\overline{\lambda}}{|\lambda|^2 - 1}\left(X^2 -\left(\lambda    
                         - \frac{1}{\overline{\lambda}}\right)X -
           \frac{\lambda}{\overline{\lambda}}\mathbf{1}\right) = 0 
\end{equation}
by the Cayley-Hamilton theorem.  Thus $B$ must still commute with $X.$ 

We now argue as in {\bf{Case 1}} of Section 5.6 and conclude that the
state must be one of the slice states \eqref{prime1} or \eqref{prime2},
but with $\lambda$ replaced by $1/\overline{\lambda}.$

\subsection{Non-generic Type 4: $T_1{T_2}^{-1}=\lambda \mathbf{1}.$}

The assumption ({\bf{Gen 4}}) stated that the only matrices that
commute with the $2\times2$ matrix $X=T_1{T_2}^{-1}$ are linear
combinations of $\mathbf{1}$ and $X$ itself.  This fails only if
$X$ is a multiple of the identity, in which case $T_2 = \lambda T_1$ and
the state is factorisable:
\begin{equation}
 |\Psi \rangle = \left( |\up \rangle + \lambda |\dn \rangle
\right)\sum_{i,j} t_{1ij} |\psi_i \rangle |\psi_j \rangle,
\end{equation} 
so that particle $1$ is a bystander.

\subsection{Non-generic Type 5} 

The assumption ({\bf{Gen 5}}) was the statement that $\alpha
\mathbf{1} + \beta T_1 {T_2}^{-1}$ is not hermitian unless $\beta=0.$
Suppose this is not true, i.e., 
\begin{equation}\label{CHTT}
 T_1 {T_2}^{-1} = u \mathbf{1} + vB
\end{equation}
where $u$ and $v$ are complex scalars and $B$ is hermitian and
traceless.  To analyse states of this form, let us assume that the basis
states of particle $1$ have been chosen by means of a Schmidt
decomposition of the three-particle state $|\Psi \rangle,$ so that the
two-particle states
\begin{equation}
 | \Phi_1 \rangle = \sum_{i,j} t_{1ij} |\psi_i \rangle |\psi_j \rangle 
\end{equation}
and
\begin{equation}
 |\Phi_2 \rangle = \sum_{i,j} t_{2ij} |\psi_i \rangle |\psi_j \rangle
\end{equation}
are orthogonal.  Let us also suppose that the basis states of particles
$2$ and $3$ have been chosen so that $T_2$ is diagonal.  Writing
\begin{align}
 &T_2 = 
 \begin{pmatrix} p & 0 \\
                 0 & q 
 \end{pmatrix}, \quad
 &B = 
 \begin{pmatrix} r & z \\
                 \overline{z} & -r
 \end{pmatrix},
\end{align}
we then have 
\begin{equation}
 T_1 =
 \begin{pmatrix} p(u+rv) & qzv \\
                 p\overline{z}v & q(u-rv) 
 \end{pmatrix}
\end{equation}
and the orthogonality of $|\Phi_1 \rangle$ and $|\Phi_2 \rangle$ gives
\begin{equation}\label{orthcond}
 p^2(u+rv) + q^2(u-rv) = 0.
\end{equation}

Now from \eqref{BTTC2} and the following line, the traceless hermitian
matrix $C$ is given by
\begin{equation}
 C\trans = -{T_2}^{-1} B T_2 =
 \begin{pmatrix} -r & -p^{-1}qz \\
                 -q^{-1}p\overline{z} & r
 \end{pmatrix}.
\end{equation}
Since this is hermitian and $p$ and $q$ are real, $p^2=q^2.$  Now
\eqref{orthcond} gives us $u=0,$ so the state is
\begin{multline}
 |\Psi \rangle = p |\dn \rangle \left(|\up\rangle |\up
       \rangle \pm |\dn \rangle |\dn \rangle \right) \\ 
       + pv |\up\rangle \left[r\left(|\up \rangle |\up
\rangle \mp |\dn \rangle |\dn \rangle \right) \pm z
|\up \rangle |\dn \rangle + \overline{z} |\dn \rangle
|\up \rangle \right]\notag.
\end{multline}
We can choose the upper sign (the state with the lower sign is related
to it by changing the sign of $|\dn_3\rangle$).  Then $T_2$ is a
multiple of the identity and $T_1$ is hermitian, so both $T$-matrices
can be simultaneously diagonalised.  Since tr$T_1 = 0,$ this gives a
state of the form 
\begin{equation}
 |\Psi\rangle = \frac{1}{\sqrt{2}} \cos\alpha |\dn\rangle
  \left(|\up\rangle|\up\rangle + |\dn\rangle|\dn\rangle\right) +
  \frac{1}{\sqrt{2}} \sin\alpha \left( |\up\rangle|\up\rangle -
  |\dn\rangle|\dn\rangle\right)\notag . 
\end{equation}
Relabelling particles $1$ and $2$ gives
\begin{align}
 |\Psi\rangle &= \frac{1}{\sqrt{2}} |\up\rangle
  \left(\cos\alpha|\dn\rangle|\up\rangle + \sin\alpha
  |\up\rangle|\up\rangle \right)\notag \\
  &+ \frac{1}{\sqrt{2}} |\dn\rangle
  \left(\cos\alpha |\dn\rangle|\dn\rangle - \sin\alpha
  |\up\rangle|\dn\rangle \right)\notag \\
  &= \frac{1}{\sqrt{2}} |\up\rangle|\up'\rangle|\up\rangle +
\frac{1}{\sqrt{2}} |\dn\rangle \left(\cos 2\alpha |\up'\rangle|\dn\rangle
- \sin 2\alpha |\dn'\rangle|\dn\rangle\right)\notag 
\end{align}
where $|\up'\rangle = \cos\alpha |\dn\rangle + \sin\alpha |\up\rangle$ and
$|\dn'\rangle = - \sin\alpha |\dn\rangle + \cos \alpha |\up\rangle.$ 
This is a slice ridge state.

This completes the classification theorem. $\blacksquare$

\section{A bestiary of atypical pure states of three spin-$\tfrac{1}{2}$ particles.}

In this section we will summarise the findings of the previous section by
describing all pure three-particle states with exceptional types of
entanglement.  We will describe their place in the space of 
all pure three-particle states, using the
canonical form of Linden, Popescu and Schlienz (henceforth called the LPS
normal form) from \cite{LnP,Schlienz}.  These authors pointed out that
any normalised three-particle state can be brought by local unitary
operations to the form
\begin{multline}
\cos \alpha |\up \rangle \left( \cos \beta |\up \rangle |\up \rangle
            + \sin \beta |\dn \rangle |\dn \rangle \right) \\
+ \sin \alpha |\dn \rangle \left(-t \sin \beta |\up \rangle
|\up \rangle + t \cos \beta |\dn \rangle |\dn \rangle
+ s|\up \rangle |\dn \rangle + z |\dn \rangle
|\up \rangle \right) 
\end{multline}
where $\alpha$ and $\beta$ are angles lying between $0$ and
$\tfrac{\pi}{4}, \quad t$ and $s$ are real and positive, and 
\begin{equation}
 s^2 + t^2 + |z|^2 = 1.
\end{equation}
In accord with our remark at the end of section 3, there are five
independent parameters (the sixth being the norm which we are taking to
be 1).  States with different values of these five parameters are
locally inequivalent, except that when $r=0$ or $s=0$ we may change the
phase of $z,$ which may therefore be taken to be real and positive; and
when $\alpha=0$ all values of $(s,t,z)$ give the same state.

We will also give an indication of the exceptional nature of these
states and their physical significance by calculating their 2-tangles
and 3-tangles. These invariants, which were introduced by Wootters
\cite{Wootquant}, quantify how much of the entanglement is contained 
in particular pairs and how much is an essential property of the full
set of three particles. Formulae for them were given by Coffman, Kundu
and Wootters \cite{tangle}. For a pure three-particle state, the
2-tangle of particles $A$ and $B$ is 
\begin{equation}
  \tau_{AB} = [\max \{\lambda_1 - \lambda_2 - \lambda_3 - \lambda_4, 0 \}]^2
\end{equation}
where $(\lambda_1, \lambda_2, \lambda_3, \lambda_4)$ are, in decreasing
order of magnitude, the positive square roots of the eigenvalues of 
\begin{equation}
  \rho_{AB}\widetilde{\rho}_{AB} = \rho_{AB}(\rho_{AB} - \rho_A - \rho_B + 1),
\end{equation}
$\rho_{AB}$ being the reduced density matrix of the pair $(A,B)$,
obtained from $|\Psi\rangle\langle \Psi|$ by tracing over particle $C$, while $\rho_A,
\rho_B$ are the reduced density matrices of particles $A$ and $B$. The
3-tangle is 
\begin{equation}
  \tau_{ABC} = 4\det\rho_A - \tau_{AB} - \tau_{AC}
\end{equation}
which can be shown \cite{tangle} to be invariant under permutations of
$A$, $B$ and $C$.

The exceptional states are as follows.

\subsection{Bystander States}

These are states which factorise as the product of a one-particle state and a
two-particle state, so that the one particle is a bystander.  They occur
when the LPS parameters have the values $\alpha = 0$ or
$\beta=0, \quad s=t=0$ or $\beta=0, \quad s=z=0.$  The state given by
$\alpha=0,$ namely
\begin{equation*}
 |\up\rangle \left(\cos \beta |\up\rangle|\up\rangle +
           \sin \beta |\dn\rangle |\dn \rangle \right)
\end{equation*}
has the two-dimensional stabiliser
\begin{equation*}
 \left(e^{i\varphi},U,V,W \right) =
 \left(e^{i\theta}, e^{-i\theta \sigma_3}, e^{i\kappa \sigma_3},
                    e^{-i\kappa \sigma_3} \right)
\end{equation*}
unless $\beta=\tfrac{\pi}{4}$ when the two-particle state is maximally
entangled and the stabiliser is four-dimensional:
\begin{equation*}
 \left(e^{i\varphi},U,V,W \right) =
 \left(e^{i\theta},e^{-i\theta \sigma_3}, V, \overline{V} \right)
\end{equation*}
or $\beta=0,$ when the state is completely factorisable and the
stabiliser is three-dimensional:
\begin{equation*}
 \left(e^{i\varphi},U,V,W \right) =
 \left(e^{i\varphi}, e^{i\theta \sigma_3}, e^{i\kappa \sigma_3},
                     e^{i\eta \sigma_3} \right)
\end{equation*}
with $\varphi + \theta + \kappa + \eta = 0.$

The 2-tangles and 3-tangle of this state are
\[
  \tau_{12} = \tau_{13} = 0, \qquad \tau_{23} = \sin^2 2\beta ,
\]
\[
  \tau_{123} = 0.
\]

\subsubsection{The General Slice State}

These are states given by
\begin{align*}
   &T_1 = 
   \begin{pmatrix} p & 0 \\
                   0 & 0
   \end{pmatrix},
   &T_2 =
   \begin{pmatrix} 0 & 0 \\
                   r & q 
   \end{pmatrix}
\end{align*}
and their relatives obtainable by permuting the particles: explicitly,    
\begin{align*}
 &p|\up \up \up \rangle + q| \dn \dn
\dn \rangle + r|\dn \dn \up \rangle, \\
 &p|\up \up \up \rangle + q| \dn \dn
\dn \rangle + r|\dn \up \dn \rangle, \\
 &p|\up \up \up \rangle + q| \dn \dn
\dn \rangle + r|\up \dn \dn \rangle.
\end{align*}
Such states occur among
the LPS normal forms when $\alpha \neq 0$ and any two of $\beta,s$ and
$z$ are zero. They have one-dimensional stabilisers each consisting of
four circles; for the first state listed above,
the stabiliser contains
\begin{equation}\label{slicesum}
 (e^{i\varphi},U,V,W) = \left( \varepsilon_1, e^{i\theta
\sigma_3}, \varepsilon_2 e^{-i\theta \sigma_3}, \varepsilon_1
\varepsilon_2 \mathbf{1} \right).              
\end{equation}
where $\varepsilon_1, \varepsilon_2 = \pm 1.$  Its tangle invariants are
\begin{equation} \label{slicetangle2}
  \tau_{12} = 4|p|^2 |r|^2, \qquad \tau_{13} = \tau_{23} = 0,
\end{equation} 
\begin{equation} \label{slicetangle3}
  \tau_{123} = 4|p|^2 |q|^2.
\end{equation}

\subsection{The Maximal Slice State, or ``Slice Ridge''}

These states, which are those slice states that have maximal values of
two out of the three two-particle von Neumann entropies, occur when a
Slice state has $|p|^2 = |q|^2 + |r|^2 = 1/2,$ i.e.
$\alpha=\tfrac{\pi}{4}$ in the Linden-Popescu normal form. In
addition to the other slice stabiliser elements \eqref{slicesum}, they
have a further one-dimensional set of stabiliser elements given, for
states in LPS normal form with $\beta=0,\,s=0,\,t=\cos\gamma$ and
$z=\sin\gamma,$ by  
\begin{multline*}
 \left(e^{i\varphi},U,V,W \right) = \\
 \left( \varepsilon_1 i,
 \begin{pmatrix} 0 & e^{i\theta} \\
                 -e^{-i\theta} & 0 
 \end{pmatrix},
 \varepsilon_2
 \begin{pmatrix} 0 & e^{-i\theta} \\
                 -e^{i\theta} & 0 
 \end{pmatrix},
 -i\varepsilon_1 \varepsilon_2 
 \begin{pmatrix} \sin\gamma & \cos\gamma\\
                 \cos\gamma & -\sin\gamma
 \end{pmatrix}
 \right) 
\end{multline*}
where $\varepsilon_1, \varepsilon_2 = \pm 1$ and $\theta$ can take any value between $0$ and
$2\pi.$  

The tangles of these states continue to be given by \eqref{slicetangle2}
and \eqref{slicetangle3}. Note that for given $p$, the maximum 3-tangle
occurs at $r=0$, when the state belongs to the following class and the
stabiliser becomes two-dimensional.

\subsection{Generalised GHZ States}

Occurring at the boundary of the set of slice states, these states are
of the form  
\begin{align*}
 &p|\up \up \up \rangle + q |\dn \dn
\dn \rangle \quad &(|p| \neq |q|).
\end{align*}
They have two-dimensional stabilisers
\begin{equation}\label{ghzgen}
 \left(e^{i\varphi},U,V,W \right)=
  \left(\pm \mathbf{1}, e^{i\theta \sigma_3}, e^{i\kappa \sigma_3},           
   e^{i\eta \sigma_3} \right)               
\end{equation}
with $\theta + \kappa + \eta = 0$ or $\pi.$  In LPS normal
form, these states have $\beta=0,s=0$ and $z=0.$ These states have pure
three-particle entanglement, since each of their two-particle density
matrices is 
\[
  \rho_{12} = \rho_{13} = \rho_{23} = |p|^2|\up\up\rangle\langle\up\up|
   + |q|^2|\dn\dn\rangle\langle\dn\dn| 
\]
which is separable. This is shown by the tangle invariants:
\[
  \tau_{12} = \tau_{13} = \tau_{23} = 0,
\]
\begin{equation} \label{GHZtangle}
  \tau_{123} = 4|p|^2|q|^2.
\end{equation}

\subsection{The true GHZ State} 

This occupies the same position among the generalised GHZ states as the
slice ridge states among the general slice states, occurring when
$|p|=|q|$ ($\alpha=\tfrac{\pi}{4}$ in LPS normal form), which maximises
the 3-tangle \eqref{GHZtangle}.  In
addition to the stabiliser elements \eqref{ghzgen}, it has the further
two-dimensional set of stabiliser elements
\begin{equation*}
 \left(e^{i\varphi},U,V,W \right) =
  \left(\pm i, i \sigma_2 e^{i \theta \sigma_3}, i \sigma_2
               e^{i\kappa\sigma_3}, i \sigma_2 e^{i\eta \sigma_3}\right)               
\end{equation*}
with $\theta + \kappa + \eta = 0.$

\subsection{The Singular Tetrahedral, or ``Beechnut'' State}

We call ``tetrahedral" states of the form 
\begin{align*}
   &T_1 = 
   \begin{pmatrix} s & 0 \\
                   0 & p
   \end{pmatrix},
   &T_2 =
   \begin{pmatrix} 0 & q \\
                   r & 0 
   \end{pmatrix}
\end{align*}
since when the eight coefficients $t_{ijk}$ are laid out in a $2\times
2$ cubic array, these states have zero entries except at the vertices of
a tetrahedron. If all four of $a,b,c,d$ are non-zero, the state is
generic. If one of them is zero, say $s=0$, the state is of the form
\[
  p|\up\dn\dn\rangle + q|\dn\up\dn\rangle + r|\dn\dn\up\rangle
\]
which has the one-dimensional stabiliser
\begin{equation*}
  \left(e^{i\varphi},U,V,W \right) =
  \left(e^{i\varphi}, e^{i\varphi \sigma_3}, e^{i\varphi \sigma_3},
                      e^{i\varphi\sigma_3} \right)
\end{equation*}
Its tangle invariants are
\begin{align*}
  \tau_{12} &= 4|p|^2|q|^2 \\
  \tau_{13} &= 4|p|^2|r|^2 \\
  \tau_{23} &= 4|q|^2|r|^2,
\end{align*}
\[
 \tau_{123} = 0.
\]

These states are, in a sense, the opposites of the generalised GHZ
states: their entanglement is concentrated in two-particle entanglement,
and they have no three-particle entanglement.

\section{Conclusion}

We have mapped the full range of entanglement properties of pure states
of three spin-$\tfrac{1}{2}$ particles, using their behaviour under
local unitary transformations as an indicator.  We have identified all
the types of exceptional states, and have shown that these states will
have a special relation to certain local invariants.  In future work we
hope to identify these invariants, and to study more fully the variation
of known invariants, such as the two-particle von Neumann entropies,
with respect to entanglement type.

\newpage

\section{Appendix: The Bestiary's Family Album}

\begin{figure}[h]
 \begin{center}
 \includegraphics{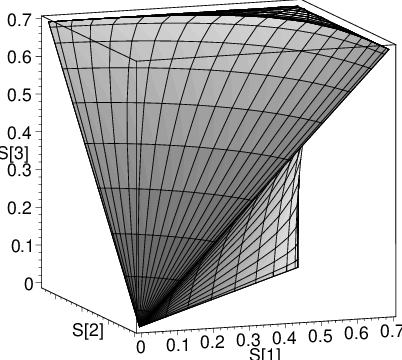}
 \includegraphics{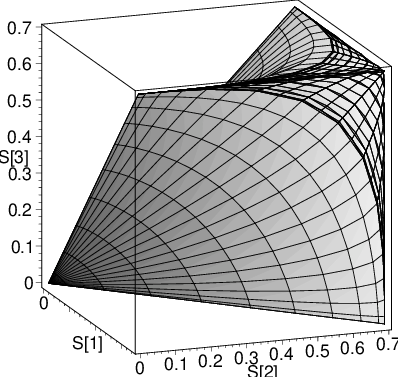}
 \caption{{\bf{The Pod: }} \emph{Here is the space of all possible pure states
of three spin-$\tfrac{1}{2}$ particles, shown from two angles.  The ``hiccup''
or seam in the parametrisation lines is not a graphical artefact, it is the
line where the pod surface ceases to be identical to the beechnut
surface (see figure \ref{beechpic}).}}\label{podpic}
 \end{center}
\end{figure}

\begin{figure}
 \begin{center}
 \includegraphics{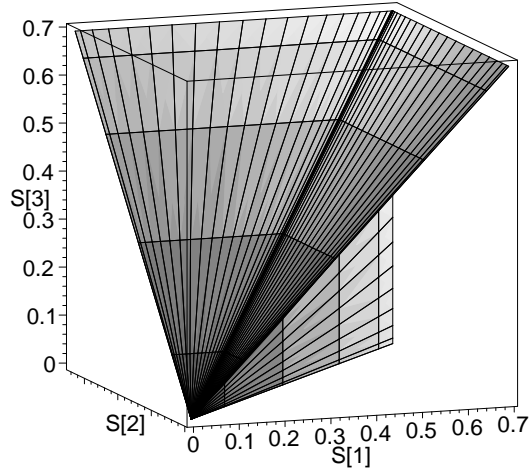}
 \caption{{\bf{The Slice States: }} \emph{The von Neumann entropies for
 all three sets of slice states.  The central spine linking all three fins is the
 subspace of generalised GHZ states, with the maximally entangled GHZ
 state at the top end, and the spin eigenstate at the bottom. 
 The outside corners are the three possible two-particle maximally
entangled states
 (with the other particle a bystander), and the edges running from those
 corners to the spin eigenstate have non-maximal two-particle entanglement,
 but are still bystander states.  The edges that run from the points of
 each maximal two-particle entanglement to the maximal GHZ state are the slice
 ridges.}}\label{slicepic} 
 \end{center}
\end{figure}

\begin{figure}
 \begin{center}
 \includegraphics{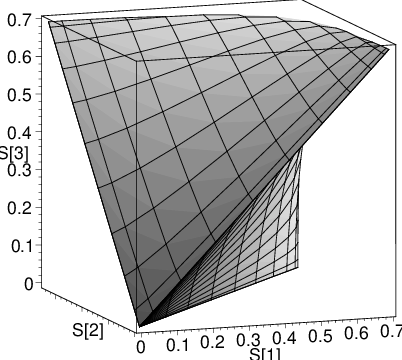}
 \includegraphics{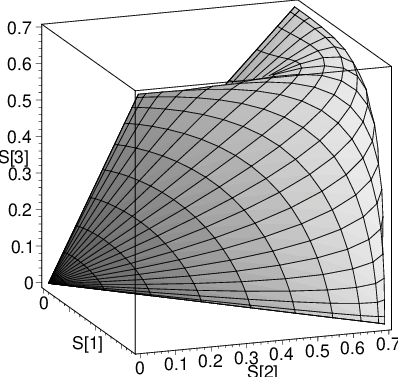}
 \caption{{\bf{The Beechnut: }}\emph{Here's how the Beechnut states got
their name.  These are the same graph, seen from two
angles.  Note that the dome at the top of the Beechnut doesn't reach the maximally
entangled GHZ state.}}\label{beechpic}
 \end{center}
\end{figure}

In this collection of figures we reproduce some graphs of the two-particle
subsystem von Neumann entropies for the various kinds of non-generic
state.  First of all, let us look at the space of all possible pure states
of three spin-$\tfrac{1}{2}$ particles, a shape we nicknamed ``The Pod''
in figure \ref{podpic}.  Then there are the Slice States in figure
\ref{slicepic} and the Beechnut states in figure \ref{beechpic}.

\end{document}